\documentclass[twocolumn,aps,prd,superscriptaddress,showpacs]{revtex4}

\usepackage{graphicx}
\usepackage{dcolumn}
\usepackage{amsmath}
\usepackage{epsfig}
\usepackage{relsize}
\RequirePackage{xspace}

\input{babarsym.tex}

\def\etal     {{\it et\,al.}}

\def\babar{\mbox{\slshape B\kern-0.1em{\smaller A}\kern-0.1em B\kern-0.1em{\smaller A\kern-0.2em R}}}

\def\invfb    {fb\ensuremath{^{-1}}\xspace}
\def\Ds       {\ensuremath{D^-_s}\xspace}
\def\Dsm       {\ensuremath{D^-_s}\xspace}
\def\Dsp       {\ensuremath{D^+_s}\xspace}
\def\Dss      {\ensuremath{D^{*-}_s}\xspace}
\def\Dssm      {\ensuremath{D^{*-}_s}\xspace}
\def\Dz       {\ensuremath{D^0}\xspace}
\def\Dbar     {\kern 0.2em\overline{\kern -0.2em D}{}\xspace}
\def\Dzb      {\ensuremath{\bar{D}^0}\xspace}
\def\Dp       {\ensuremath{D^+}\xspace}
\def\Dm       {\ensuremath{D^-}\xspace}

\def\taup     {\ensuremath{\tau^+}\xspace}
\def\taum     {\ensuremath{\tau^-}\xspace}

\def\ep       {\ensuremath{e^+}\xspace}
\def\en       {\ensuremath{e^-}\xspace}
\def\epem     {\ensuremath{e^+e^-}\xspace}
\def\nub      {\ensuremath{\bar{\nu}}\xspace}

\def\nuellb      {\ensuremath{\nub_\ell}\xspace}
\def\nue      {\ensuremath{\nu_e}\xspace}
\def\nueb      {\ensuremath{\nub_e}\xspace}

\def\numub     {\ensuremath{\nub_{\mu}}\xspace}
\def\nutau      {\ensuremath{\nu_\tau}\xspace}
\def\nutaub     {\ensuremath{\nub_\tau}\xspace}

\def\cbar       {\ensuremath{\bar c}\xspace}
\def\ccbar    {\ensuremath{c\bar c}\xspace}
\def\s        {\ensuremath{s}\xspace}

\def\Wm       {\ensuremath{W^-}\xspace}

\def\Vcs      {\ensuremath{|V_{cs}|}\xspace}

\def\mup      {\ensuremath{\mu^+}\xspace}
\def\mum      {\ensuremath{\mu^-}\xspace}
\def\piz      {\ensuremath{\pi^0}\xspace}
\def\pip      {\ensuremath{\pi^+}\xspace}
\def\pim      {\ensuremath{\pi^-}\xspace}

\def\KL       {\ensuremath{K^0_{\scriptscriptstyle L}}\xspace}
\def\KS       {\ensuremath{K^0_{\scriptscriptstyle S}}\xspace}
\def\Kp       {\ensuremath{K^+}\xspace}
\def\Km       {\ensuremath{K^-}\xspace}
\def\Kpm      {\ensuremath{K^\pm}\xspace}
\def\kaon     {\ensuremath{K}\xspace}
\def\Kbar     {\kern 0.2em\overline{\kern -0.2em K}{}\xspace}
\def\BR       {{\ensuremath{\cal B}\xspace}}
\def\g        {\ensuremath{\gamma}\xspace}

\def\mDs{\ensuremath{m_{r}(DKX\gamma)}}
\def\mmmunu{\ensuremath{m^2_{r}(DKX\gamma\mu)}}
\def\mmenu{\ensuremath{m^2_{r}(DKX\gamma e)}}
\def\mKKpig{\ensuremath{m(KK\pi\gamma)}}
\def\fDs{\ensuremath{f_{D_s}}}

\def\ellm      {\ensuremath{\ell^-}\xspace}

\def\DsToEllNu{\ensuremath{\Dsm\!\rightarrow\!\ellm\nub_{\ell}}\xspace}
\def\DsToMuNu{\ensuremath{\Dsm\!\rightarrow\!\mum\numub}\xspace}
\def\DsToENu{\ensuremath{\Dsm\!\rightarrow\!\en\nueb}\xspace}
\def\DsToTauNu{\ensuremath{\Dsm\!\rightarrow\!\taum\nutaub}\xspace}
\def\TauToMuNuNu{\ensuremath{\taum\!\rightarrow\!\mum\numub\nutau}\xspace}
\def\TauToENuNu{\ensuremath{\taum\!\rightarrow\!\en\nueb\nutau}\xspace}
\def\DsToTauNuENuNu{\ensuremath{\Dsm\!\rightarrow\!\taum_{e\nu\nu}\nutaub}\xspace}
\def\DsToTauNuMuNuNu{\ensuremath{\Dsm\!\rightarrow\!\taum_{\mu\nu\nu}\nutaub}\xspace}

\def\dEdx  {\ensuremath{dE/dx}}
\def\mevcc  {MeV\xspace}
\def\mevc  {MeV\ensuremath{/c}\xspace}
\def\mevcc  {MeV\ensuremath{/c^2}\xspace}
\def\gev  {GeV\xspace}
\def\gevc  {GeV\ensuremath{/c}\xspace}
\def\gevcc  {GeV\ensuremath{/c^2}\xspace}
\def\GeVccSq  {GeV\ensuremath{^2/c^4}}
\def\Eextra  {\ensuremath{E_{extra}}}

\def\NDs{\ensuremath{N_{D_s}}}
\def\Nmunu{\ensuremath{N_{\mu\nu}}}
\def\Nenu{\ensuremath{N_{e\nu}}}
\def\Nmunu{\ensuremath{N_{\mu\nu}}}

\def\NKKpi{\ensuremath{N_{KK\pi}}}

\def\epsmunu{\ensuremath{\bar{\varepsilon}_{\mu\nu}}}
\def\epsenu{\ensuremath{\bar{\varepsilon}_{e\nu}}}

\def\epsKKpi{\ensuremath{\bar{\varepsilon}_{KK\pi}}}

\def\Belle{Belle}

\newcommand{\BABARPubYear}    {10}
\newcommand{\BABARPubNumber}  {024}
\newcommand{\SLACPubNumber} {14229}
\newcommand{\LANLNumber} {1008.4080}

\def\fDsValue{258.6} 
\def\fDsValueStat{6.4} 
\def\fDsValueSyst{7.5} 
\def\nfDsSD{1.8}

\def\figurebox#1#2#3{
    \def\arg{#3}
    \ifx\arg\empty
    {\hfill\vbox{\hsize#2\hrule\hbox to #2{\vrule\hfill\vbox to #1{\hsize#2\vfill}\vrule}\hrule}\hfill}
    \else
    {\hfill\epsfbox{#3}\hfill}
    \fi}

\begin{document}

\preprint{\babar-PUB-\BABARPubYear/\BABARPubNumber}
\preprint{SLAC-PUB-\SLACPubNumber}

\begin{flushleft}
\ \\
\ \\
\end{flushleft}

\

\

\begin{flushright}
  \babar-PUB-\BABARPubYear/\BABARPubNumber\\
  SLAC-PUB-\SLACPubNumber\\
  hep-ex/\LANLNumber\\
\end{flushright}

\title{
{\large {\boldmath
Measurement of the Absolute Branching Fractions for $D^-_s\rightarrow\ell^-\bar{\nu}_{\ell}$ and Extraction of the Decay Constant $f_{D_s}$
}}
}

\date{\today}

%
\author{P.~del~Amo~Sanchez}
\author{J.~P.~Lees}
\author{V.~Poireau}
\author{E.~Prencipe}
\author{V.~Tisserand}
\affiliation{Laboratoire d'Annecy-le-Vieux de Physique des Particules (LAPP), Universit\'e de Savoie, CNRS/IN2P3,  F-74941 Annecy-Le-Vieux, France}
\author{J.~Garra~Tico}
\author{E.~Grauges}
\affiliation{Universitat de Barcelona, Facultat de Fisica, Departament ECM, E-08028 Barcelona, Spain }
\author{M.~Martinelli$^{ab}$}
\author{A.~Palano$^{ab}$ }
\author{M.~Pappagallo$^{ab}$ }
\affiliation{INFN Sezione di Bari$^{a}$; Dipartimento di Fisica, Universit\`a di Bari$^{b}$, I-70126 Bari, Italy }
% ...
\collaboration{The \babar\ Collaboration}
\noaffiliation

\begin{abstract}

The absolute branching fractions for the decays $D^-_s\!\rightarrow\!\ell^-\bar{\nu}_{\ell}$ ($\ell=e$, $\mu$, or $\tau$) are measured using a data sample corresponding to an integrated luminosity of 521 fb$^{-1}$ collected at center of mass energies near 10.58 GeV with the \mbox{\slshape B\kern-0.1em{\smaller A}\kern-0.1em B\kern-0.1em{\smaller A\kern-0.2em R}} detector at the PEP-II $e^+e^-$ collider at SLAC.
The number of $\Dsm$ mesons is determined by reconstructing the recoiling system $DKX\gamma$ in events of the type $e^+e^-{\rightarrow}DKXD^{*-}_s$, where $D^{*-}_s\rightarrow D^-_s\gamma$ and $X$ represents additional pions from fragmentation. 
The $D^-_s\rightarrow\ell^-\nu_{\ell}$ events are detected by full or partial reconstruction of the recoiling system $DKX\gamma\ell$.
The branching fraction measurements are combined to determine the $D^-_s$ decay constant $f_{D_s} = (258.6 \pm 6.4 \pm 7.5)$ MeV, where the first uncertainty is statistical and the second is systematic.

\end{abstract}

\pacs{13.20.Fc,12.38.Gc}

\maketitle

The $\Dsm$ meson can decay purely leptonically
via annihilation of the $\cbar$ and $\s$ quarks into a $\Wm$ boson~\cite{ref:chargeconj}.
In the Standard Model (SM), the leptonic partial width $\Gamma(\DsToEllNu)$ is given by 
\begin{equation}
   \Gamma= 
   \frac{G^2_F M_{D_s}^3 }{8\pi}  
   \left(\frac{m_{\ell}}{M_{D_s}}\right)^2
   \left(1-\frac{m_{\ell}^2}{M_{D_s}^2}\right)^2 
   \Vcs^2f^2_{D_s},\\
\label{eq:Gamma}
\end{equation}
where $M_{D_s}$ and $m_{\ell}$ are the $\Dsm$ and lepton masses, respectively,
$G_F$ is the Fermi coupling constant, and
$V_{cs}$ is an element of the Cabibbo-Kobayashi-Maskawa quark mixing matrix. 
These decays provide a clean probe of the pseudoscalar meson decay constant $f_{D_s}$.

Within the SM, \fDs has been predicted using several methods \cite{ref:uqlqcd}; the most precise value by Follana \etal~ uses unquenched LQCD calculations and gives \fDs=$(241\pm3)$ \mev.
Currently, the experimental values are significantly larger than this theoretical prediction.
The Heavy Flavor Averaging Group combines the CLEO-c, \Belle~ and \babar~ measurements and reports \fDs = (254.6 $\pm$ 5.9) \mev \cite{ref:HFAG}.
Models of new physics (NP), including a two-Higgs doublet \cite{ref:twoHiggs} and leptoquarks \cite{ref:newPhysics}, may  explain this difference.
In addition, \fDs measurements provide a cross-check of QCD calculations which predict the impact of NP on $B$ and $B_s$ meson decay rates and mixing.
High precision determinations of $\fDs$, both from experiment and theory, are necessary in order to discover or constrain effects of NP.

We present absolute measurements of the branching fractions of leptonic \Ds decays with a method similar to the one used by the \Belle~ Collaboration~\cite{ref:belleMuNu,ref:belleSL}.
An inclusive sample of $\Ds$'s is obtained by reconstructing the rest of the event in reactions of the kind \mbox{$\ep\en\!\rightarrow\!\ccbar\!\rightarrow\! D K X \Dssm$}, where $\Dssm\!\rightarrow\!\Dsm\g$.
Here, $D$ represents a charmed hadron ($D^0$, $D^+$, $D^*$, or $\Lambda_c^+$), $K$ represents the $\KS$ or $\Kp$ required to balance strangeness in the event, and $X$ represents additional pions produced in the $\ccbar$ fragmentation process. 
When the charmed hadron is a $\Lambda_c^+$ an additional anti-proton is required to assure baryon number conservation.
No requirements are placed on the decay products of the \Ds~ so that the selected events correspond to an inclusive sample.
The 4-momentum of each $\Dsm$ candidate, $p_r$, is measured as the difference between the momenta of the colliding beam particles and the fully reconstructed $DKX\gamma$ system:~$p_{r}~=~p_{e^+}~+~p_{e^-}~-~p_{D}~-~p_{K}~-~p_{X}~-~p_{\gamma}$.
The inclusive $\Dsm$ yield is obtained from a binned fit to the distribution in the recoil mass $\mDs\equiv\sqrt{p^2_{r}}$.
Within this inclusive sample, we determine the fraction of events corresponding to \DsToMuNu, \DsToENu, and \DsToTauNu decays.
In the SM, ratios of the branching fractions for these decays are $\en\nueb$:$\mum\numub$:$\taum\nutaub$=$2\times10^{-5}$: 1 : 10, due to helicity and phase-space suppression.

The analysis is based on a data sample of 521 \invfb, which corresponds to about 677 million $\epem\to\ccbar$ events, recorded near $\sqrt{s}=10.58$ GeV by the $\babar$ detector at the SLAC \pep2 asymmetric-energy collider.
The detector is described in detail in Refs.~\cite{ref:babar,ref:babar2}.
Charged-particle momenta are measured with a 5 layer, double-sided silicon vertex tracker (SVT)
and a 40 layer drift chamber (DCH) inside a 1.5 T superconducting solenoidal magnet.
A calorimeter consisting of 6580 CsI(Tl) crystals (EMC) is used to measure
electromagnetic energy. Measurements from a ring-imaging Cherenkov radiation detector, and of specific ionization (\dEdx) in the SVT and DCH, provide particle identification (PID) of charged hadrons. Muons are mainly identified
by the instrumented magnetic flux return, and electrons are identified using EMC and \dEdx~ information.
The analysis uses Monte Carlo (MC) events generated with EvtGen and JETSET \cite{ref:evtgen,ref:jetset} and passed through a detailed GEANT4 \cite{ref:geant4} simulation of the detector response. Final state radiation from charged particles is modeled by PHOTOS~\cite{ref:photos}.
Samples of MC events for \epem annihilation to $q\bar{q}$ ($q=u,d,s,c,b$) (generic MC) are used to develop methods to separate signal events from backgrounds. In addition, we use dedicated samples for \Dsm production and leptonic decays (signal MC) to determine reconstruction efficiencies and the distributions needed for the extraction of the signal decays.

We reconstruct $D$ candidates using the following 15 modes:
{$\Dz\rightarrow\Km\pip(\piz)$, $\Km\pip\pim\pip(\piz)$, or $\KS\pip\pim(\piz)$;}
{$\Dp\rightarrow\Km\pip\pip(\piz)$, $\KS\pip(\piz)$, or $\KS\pip\pim\pip$;}
and {$\Lambda_c^+\to p\Km\pip(\piz)$, $p\KS$, or $p\KS\pim\pip$}. 
All $\piz$'s and \KS's used in this analysis are reconstructed from two photons or two oppositely charged pions, respectively,
and are kinematically constrained to their nominal mass values \cite{ref:pdg2008}.  
The \KS in a $D$ candidate must have a flight distance from the \epem interaction point (IP) greater than 10 times its uncertainty. 
For each $D$ candidate we fit the tracks to a common vertex, and for each mode, we determine the mean and $\sigma$ of the reconstructed signal mass distribution from a fit to data. 
We then simultaneously optimize a set of selection criteria to maximize $S/\sqrt{S+B}$,
where $S$ refers to the number of $D$ candidates after subtraction of the background $B$ within a mass window defined about the signal peak. Where $B$ is estimated from the sideband regions of the mass distribution.
In addition to the size of the mass window, several other properties of the $D$ candidate are used in the optimization: 
the center-of-mass (CM) momentum of the $D$, 
PID requirements on the tracks, 
the probability of the $D$ vertex fit,
and the minimum lab energy of $\piz$ photons.
The CM momentum must be at least 2.35 \gevc in order to remove $B$ meson backgrounds.
After the optimization the relative contributions to the total signal sample are 74.0\% \Dz, 22.6\% $D^+$, and 3.4\% $\Lambda_c^+$. Multiple candidates per event are accepted.

To identify $D$ mesons originating from $D^*$ decays we reconstruct the following decays: 
$D^{*+}\to\Dz\pip$, $D^{*0}\to\Dz\piz$, $D^{*+}\to\Dp\piz$, and $D^{*0}\to\Dz\gamma$. 
The photon energy in the laboratory frame is required to exceed 30 \mev for $\piz\to\gamma\gamma$ and 250 \mev for $D^{*0}\to \Dz\gamma$ decays. The $\gamma\gamma$ invariant mass must be within 3 sigma of the \piz peak. For all $D^*$ decays, the mass difference $m(D^*)-m(D)$ is required to be within 2.5 sigma of the peak value.

A $K$ candidate is selected from tracks not overlapping with the $D$ candidate.  
PID requirements are applied to each \Kp candidate, and a \KS candidate
must have a flight distance greater than 5 times its uncertainty.

An X candidate is reconstructed from the remaining $\pi^\pm$'s and {\piz}s not overlapping with the $DK$ candidate. 
In the laboratory frame, a $\pi^\pm$ must have a momentum greater than 100 \mevc and each photon from a \piz decay must have energy greater than 100 \mev.
We reconstruct X modes without \piz's with up to three charged pions, and modes with one \piz with up to two charged pions.
The total charge of the X candidate is not checked at this stage.

Finally, we select a $\gamma$ candidate for the signal \Dss~ decay by requiring a minimum energy of 120 \mev in the laboratory frame, and an angle with respect to the direction of the $D$ candidate momentum in the CM frame greater than 90 degrees.
This photon cannot form a \piz or $\eta$ candidate when combined with any other photon in the event. 
In addition, the cluster must pass tight requirements on the shower shape in the EMC and a separation of at least 15 cm from the impact of any charged particle or the position of any other energy cluster in the EMC.

Only $DKX\gamma$ candidates with a total charge of $+1$ are selected to form a right-sign (RS) sample, from which we extract the \Ds signal yield.
The charm and strange quark content of the $DKX$ must be consistent with recoiling from a \Dsm.
The RS sample includes candidates for which consistency cannot be determined due to the presence of a \KS.
We define a wrong-sign (WS) sample with the same charge requirement above, but by requiring that the charm and strange quark content of the $DKX$ be consistent with a recoil from a \Dsp.
The WS sample contains a small fraction of signal events due mainly to $DKX$ candidates for which the total charge is misreconstructed.
The generic MC shows that the WS sample, after subtraction of the signal contribution, correctly models the backgrounds in the RS sample.

A kinematic fit to each $DKX$ candidate is performed in which the particles are required 
to originate from a common point inside the IP region,
and the $D$ mass is constrained to the nominal value \cite{ref:pdg2008}. 
The 4-momentum of the signal \Dss is extracted as the missing 4-momentum in the event.
We require that the \Dss candidate mass be within 2.5$\sigma$ of the signal peak. 
For MC signal events, the mean is found to be consistent with the nominal value and $\sigma$ varies between 37 and 64 \mevcc depending on the number of pions in $X$. 

We perform a similar kinematic fit with the signal $\gamma$ included and with the mass recoiling against the $DKX$ constrained to the nominal \Dss mass \cite{ref:pdg2008} in order to determine the \Ds 4-momentum.
We require that the \Ds CM momentum exceed 3.0 \gevc, and that its mass be greater than 1.82 \gevcc.
After the final selections, there remain on average 1.7 \Ds candidates per event, due mainly to multiple photons that can be associated with the \Dss decay.
In order to properly count events in the fits described below, we assign weight 1/n to each \Ds candidate, where n is the number of \Ds candidates in the event.

We define $n_X^R$ and $n_X^T$ to be the number of reconstructed and true pions in the $X$ system, respectively.
The efficiency for reconstructing signal events depends on $n_X^T$. However, the $n_X^T$ distribution is expected to differ from the MC simulation due to inaccurate fragmentation functions used by JETSET.  
To correct for these inaccuracies, we extract the \Ds signal yields from a fit to the two-dimensional histogram of \mDs~versus $n_X^R$. The PDF for the signal distribution is written as a weighted sum of the MC distributions for $j=n_X^T$,
\begin{equation}
\label{eq:DenomSignalPdf}
S(m,n_X^R) =\sum_{j=0}^6 w_j S_{j}(m,n_X^R).
\end{equation}
The weights $w_j$ have to be extracted from this fit.
To constrain the shape of the weights distribution, we introduce the parameterization  $w_j \propto (j-\alpha)^\beta e^{-\gamma j} $ together with the condition $\sum_j w_j=1$. 
This parametrization is motivated by the distribution of weights in the MC. 
The value $\alpha=-1.32$ is taken from a fit to MC, whereas $\beta$ and $\gamma$ are determined from the fit to data.

The RS and WS samples are fitted simultaneously to determine the background.
The fit to the WS sample uses a signal component similar to that used in the RS fit, except that due to the small signal component, the weights are fixed to the MC values and the signal yield is determined from signal MC to be 11.8\% of the RS signal yield. 
The shapes remaining after the signal component is removed from the WS sample, $B_i(m)$ ($i=n_X^R$), are used to model the RS backgrounds.
A shape correction is applied to $B_{0}$ to account for a difference observed in the MC.
We add these components with free coefficients ($b_i$) to construct the total RS background shape:
 $B(m,n_X^R)=\sum_{i=0}^3 b_i B_{i}(m)\delta(i-n_X^R)$. 
Thus in addition to $\beta$, $\gamma$, and the total signal yield, there are 3 additional free parameters $b_i(i=0,1,2)$ in the RS fit.

\begin{figure}[t]
  \begin{center}
   \includegraphics[clip,width=\columnwidth]{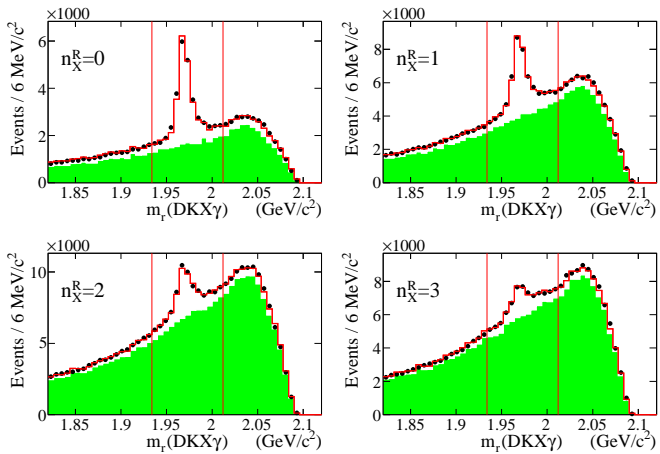}
   \caption{(color online) \mDs~ distributions for each $n_X^R$ value. The points are the data. The open histogram is from the fit described in the text. The solid histogram is the background component from the fit. The vertical lines define the region used in the $\ellm\nuellb$ selections.}
   \label{fig:DsFitVsnX}
  \end{center}
\end{figure}

\begin{figure}[t]
  \includegraphics[clip,width=\columnwidth]{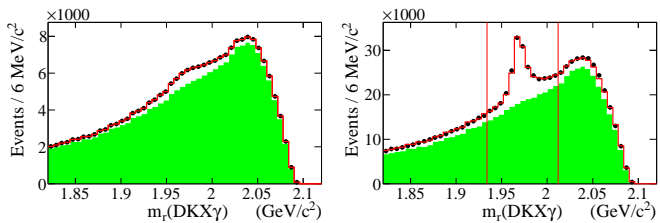}
  \caption{(color online) \mDs~ distribution for the total WS (left) and RS (right) samples.}
  \label{fig:DsFit}
\end{figure}

Figure~\ref{fig:DsFitVsnX} shows the data and the results of the fit, and Fig.~\ref{fig:DsFit} shows the total RS and WS samples. 
The fit finds a mininum $\chi^2/ndf = 216/182$ and the fitted parameter values are $\beta=0.27\pm0.17$ and $\gamma=0.28\pm0.07$. 
These are different from the MC values $\beta=3.38$ and $\gamma=1.15$ since there are more events at low values of $n_X^T$ than in the MC.

\begin{figure}[t]
  \includegraphics[clip,width=1.1\columnwidth,height=\columnwidth, angle=90]{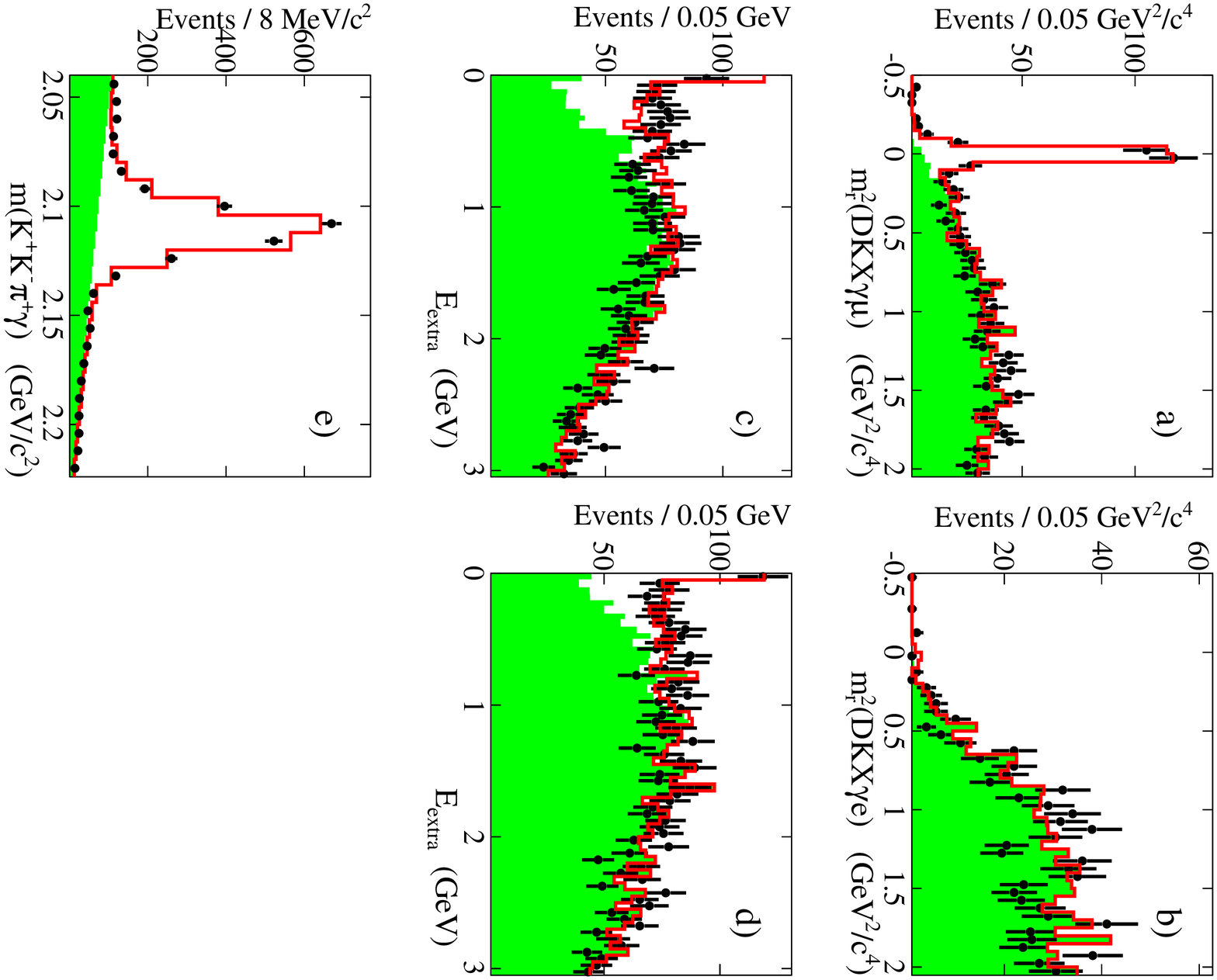}

  \caption{(color online) Fitted distributions of (a) \mmmunu, (b) \mmenu, (c) \Eextra~ for \DsToTauNuENuNu, (d) \Eextra~ for \DsToTauNuMuNuNu candidates, and (e) \mKKpig.
 In each figure, the points represent the data with statistical error bars, the open histogram is from the fit described in the text, and the solid histogram is the background component from the fit.
}
  \label{fig:NumFits}
\end{figure}

Having constructed the inclusive \Ds sample, 
we proceed to the selection of \DsToMuNu~events within that sample.
We use the \mDs~ range between 1.934 and 2.012 \gevcc, which contains an inclusive \Ds yield (\NDs) of $(67.2\pm1.5)\times 10^3$.
We require that there be exactly one more charged particle in the remainder of the event, and that it be identified as a \mum.
In addition, we require that the extra neutral energy in the event, \Eextra, be less than 1.0 \gev; 
\Eextra is defined as the total energy of EMC clusters with individual energy greater than 30 \mev and not overlapping with the $DKX\gamma$ candidate.
Since the only missing particle in the event should be the neutrino we expect the distribution of \Eextra~ to peak at zero for signal events.
We determine the 4-momentum of the $\numub$ candidate through a kinematic fit similar to that described earlier in the determination of the \Ds 4-momentum, but with the \mum included in the recoil system.
In this fit we constrain the mass recoiling against the $DKX\gamma$ system to the nominal value for the \Ds \cite{ref:pdg2008}.
To extract the signal yield, we perform a binned maximum likelihood fit to the \mmmunu~ distribution using a signal PDF determined from reconstructed signal MC events that contain the signal decay chain $\Dss\to\Ds\gamma$ with \DsToMuNu. 
The background PDF is determined from the reconstructed generic MC events with signal events removed.
The fit is shown in Fig.~\ref{fig:NumFits}(a), and the number of signal events extracted, \Nmunu, is listed in Table~\ref{tab:results}.

\begin{table*}[t] 
\caption[results]{Average efficiency ratios, signal yields, branching fractions, and decay constants for the leptonic \Ds decays. 
The first uncertainty is statistical and the second is systematic.
}  
\label{tab:results}

\begin{tabular*}{\linewidth}{
@{\extracolsep{\fill}}l
@{\extracolsep{\fill}}c
@{\extracolsep{\fill}}c
@{\extracolsep{\fill}}c
@{\extracolsep{\fill}}c
@{\extracolsep{\fill}}c
}
\hline\hline
      Decay & $\bar{\varepsilon}$ & Signal Yield & $\BR(\DsToEllNu)$ & $f_{D_s}$ (\mev)\\  
\hline \hline 
\DsToENu & 70.5\%& 6.1 $\pm$ 2.2 $\pm$ 5.2 & $<2.3\times 10^{-4}$ at 90\% C.L. & \\ 
\DsToMuNu & 67.7\% & 275 $\pm$ 17 & (6.02 $\pm$ 0.38 $\pm$ 0.34)$\times 10^{-3}$ & 265.7 $\pm$ 8.4 $\pm$ 7.7 \\ 
\DsToTauNu$(\taum\to\en\nueb\nutau)$ & 61.6\% & 408 $\pm$ 42 & (5.07 $\pm$ 0.52 $\pm$ 0.68)$\times 10^{-2}$ & 247 $\pm$ 13 $\pm$ 17\\ 
\DsToTauNu$(\taum\to\mum\numub\nutau)$ & 59.5\% & 340 $\pm$ 32 & (4.91 $\pm$ 0.47 $\pm$ 0.54)$\times 10^{-2}$ & 243 $\pm$ 12 $\pm$ 14 \\

\hline\hline
\end{tabular*} 
\end{table*}

The \DsToMuNu~ branching fraction is obtained from:
\begin{equation}
 \BR(\DsToMuNu) = { \Nmunu \over \NDs \sum_{j=0}^{6} w_j 
{\varepsilon^j_{\mu\nu} \over \varepsilon^j_{D_s} }}
= { \Nmunu \over \NDs \bar{\varepsilon}_{\mu\nu}},
 \label{eq:BFequation}
\end{equation}
where the  \DsToMuNu~ reconstruction efficiency, $\varepsilon^j_{\mu\nu}$,
is determined using the signal MC sample with $j=n_X^T$, and $\varepsilon^j_{D_s}$ 
is the corresponding inclusive \Ds reconstruction efficiency. 
The efficiency ratios ${\varepsilon^j_{\mu\nu} / \varepsilon^j_{D_s} }$ 
decrease from 87\% to 33\% as $j$ increases from 0 to 6.
The weighted average, \epsmunu, and the value determined for $\BR(\DsToMuNu)$ are listed in Table~\ref{tab:results}.
The statistical uncertainty includes contributions from 
$\NDs$, $\epsmunu$, and $\Nmunu$ (with correlations taken into accounted).
The systematic uncertainty is determined by varying the parameter values in the inclusive \Ds fit which were fixed to MC values, by varying the resolution on the \Ds signal PDF (for both mass and $n_X^R$), and by estimating how well the MC models the non-peaking component of the signal PDF observed in Figs.~\ref{fig:DsFitVsnX} and \ref{fig:DsFit}. 
The non-peaking signal component in the \mDs\ distribution arises from $DKX\gamma$ candidates in events that contain the signal decay $\Dss\to\Ds\gamma$, but for which the photon candidate is mis-identified and is due to other sources such as $\piz$ or $\eta$ decays, or tracks or \KL interacting in the calorimeter.
Uncertainties are assigned for possible mismodeling of the signal or background \mmmunu~ distributions due to possible differences in the position or resolution of the mass distribution, or mismodelings of different \Ds decays. 
Uncertainties in the efficiencies due to tracking and \mum identification are included.
This measurement supersedes our previous result \cite{ref:StelzerMuNu}.

Using a procedure similar to that for \DsToMuNu we search for \DsToENu events. 
The fit to the \mmenu~ distribution, shown in Fig.~\ref{fig:NumFits}(b), gives a signal yield  \Nenu consistent with 0.
We obtain an upper limit on $\BR(\DsToENu)$ by integrating a likelihood function from 0 to the value of $\BR(\DsToENu)$ corresponding to 90\% of the integral from 0 to infinity. 
The likelihood function consists of a Gaussian function written in terms of the variable $\BR\NDs\epsenu$ with mean and sigma set to \Nenu~ and its total uncertainty, respectively. To account for the uncertainties on $\NDs\epsenu$, the main Gaussian is convolved with another Gaussian function centered at the measured value of $\NDs\epsenu$ with sigma set to the $\NDs\epsenu$ total uncertainty.
The value obtained for the upper limit is listed in Table~\ref{tab:results}.

We find \DsToTauNu decays within the sample of inclusively reconstructed \Ds events by requiring exactly one more track identified as an \en or \mum, from the decay \TauToENuNu or \TauToMuNuNu. 
We remove events associated with \DsToMuNu decays by requiring $\mmmunu>$0.5 \GeVccSq. 
Since \DsToTauNu events contain more than one neutrino we use \Eextra~ to extract the yield of signal events; these are expected to peak towards zero, while the backgrounds extend over a wide range.
The signal and background PDFs are determined from reconstructed MC event samples.
The fits are shown in Figs.~\ref{fig:NumFits}(c) and \ref{fig:NumFits}(d); the signal yields are listed in Table~\ref{tab:results}.
We determine $\BR(\DsToTauNu)$ from the $\en$ and $\mum$ samples using Eq.~(\ref{eq:BFequation}) 
and accounting for the decay fractions of the \taum \cite{ref:pdg2008}.
The values obtained are listed in Table~\ref{tab:results} and are consistent with the previous \babar~ result~\cite{ref:shaneTauToE}. 
The error-weighted average \cite{ref:Averaging} of the branching fractions is $\BR(\DsToTauNu)=(5.00 \pm 0.35 (stat) \pm 0.49(syst))\times 10^{-2}$. The weights used in the average are computed from the total error matrix and account for correlations. As a test of lepton flavor universality we determine the ratio ${ \BR(\DsToTauNu) / \BR(\DsToMuNu)} = (8.27 \pm 0.77(stat) \pm 0.85(syst))$, which is consistent with the SM value of 9.76.

As a cross-check of this analysis method, we measure the branching fraction for the hadronic decay $\Dsm\to\Km\Kp\pim$. 
Within the inclusive \Dsm sample, we require exactly three additional charged particle tracks that do not overlap with the $DKX\gamma$ candidate. 
PID requirements are applied to the kaon candidates. 
The mass of the $\Km\Kp\pim$ system must be between 1.93 and 2.00 \gevcc, 
and the CM momentum above 3.0 \gevc. 
We combine the $\Km\Kp\pim$ system with the signal $\gamma$ and extract the signal yield from the \mKKpig~ distribution.
For this mode we choose the loose selection $\mDs>1.82$ \gevcc, because this variable is correlated with \mKKpig; this corresponds to an inclusive \Ds yield of $\NDs=(108.9\pm2.4)\times 10^3$.
We model the signal distribution using reconstructed MC events that contain the
decay chain $\Dssm\to\Dsm\gamma$ and $\Dsm\to\Km\Kp\pim$. 
In the generic MC and a high statistics control data sample (for which the inclusive reconstruction was not applied) the background was found to be linear in \mKKpig.  
From a fit to the \mKKpig~ distribution, shown in Fig.~\ref{fig:NumFits}(e), we determine a signal yield of $\NKKpi=1866 \pm 40$ events.

We compute the $\Dsm\to\Km\Kp\pim$ branching fraction using Eq.~(\ref{eq:BFequation}). 
The efficiency for reconstructing signal events is determined from the signal MC in three regions of the $\Km\Kp\pim$ Dalitz plot, corresponding to $\phi\pim$, ${\Km}K^{*0}$, and the rest.
A variation of $\sim$8\% is observed across the Dalitz plot, leading to a correction factor of 1.016 on $\varepsilon^j_{KK\pi}$.
The weighted efficiency ratio is found to be \epsKKpi=29.5\%, and we obtain $\BR(\Dsm\to\Km\Kp\pim)=(5.78\pm0.20(stat)\pm0.30(syst))\%$. 
The first uncertainty accounts for the statistical uncertainties associated with the inclusive \Dsm sample and $N_{KK\pi}$.  
The second accounts for systematic 
uncertainties in the signal and background models, and the inclusive \Dsm sample, as well as the reconstruction and PID selection of the $\Km\Kp\pim$ candidates.
This result is consistent with the value $(5.50\pm0.23\pm0.16)\%$ measured by CLEO-c \cite{ref:cleoKKpi}.

Using the leptonic branching fractions measured above, we determine the \Dsm decay constant 
using Eq.~(\ref{eq:Gamma}) and the known values for $m_{\ell}$, $m_{D_s}$, $|V_{ud}|$ (we assume $|V_{cs}|=|V_{ud}|$), and the \Ds~ lifetime obtained from Ref.~\cite{ref:pdg2008}. The \fDs~values are listed in Table~\ref{tab:results}; the systematic uncertainty includes the uncertainties on these parameters (1.9 \mev).
Finally, we obtain the error-weighted average $f_{D_s} = (\fDsValue \pm \fDsValueStat(stat) \pm \fDsValueSyst(syst))$ \mev.

In conclusion, we use the full dataset collected by the $\babar$ experiment to  measure the branching fractions for the leptonic decays of the \Dsm meson.
The measured value of \fDs~is \nfDsSD~ standard deviations larger than the theoretical value \cite{ref:uqlqcd}, consistent with the measurements by \Belle~ and CLEO-c \cite{ref:belleMuNu,ref:CLEO}. Further work on this subject is necessary to validate the theoretical calculations or to shed light on possible NP processes.

We are grateful for the excellent luminosity and machine conditions
provided by our \pep2\ colleagues, 
and for the substantial dedicated effort from
the computing organizations that support \babar.
The collaborating institutions wish to thank 
SLAC for its support and kind hospitality. 
This work is supported by
DOE
and NSF (USA),
NSERC (Canada),
CEA and
CNRS-IN2P3
(France),
BMBF and DFG
(Germany),
INFN (Italy),
FOM (The Netherlands),
NFR (Norway),
MES (Russia),
MICIIN (Spain),
STFC (United Kingdom). 
Individuals have received support from the
Marie Curie EIF (European Union),
the A.~P.~Sloan Foundation (USA)
and the Binational Science Foundation (USA-Israel).


\begin{thebibliography}{99}
\bibitem{ref:chargeconj}
  Use of charge conjugate reactions is implied in this paper.
\end{thebibliography}
\end{document}